# Nanowire Growth for Sensor Arrays

Minhee Yun[a*], Nosang V. Myung[a], Richard P. Vasquez[a], Jianjun Wang[b],
and Harold Monbouquette[b],
[a] Jet Propulsion Laboratory, California Institute of Technology, Pasadena, CA 91109
[b] Department of Chemical Engineering, University of California at Los Angeles, CA 90095

## ABSTRACT

A design concept for nanowire-based sensors and arrays is described. The fabrication technique involves electrodeposition to directly grow nanowires between patterned thin film contact electrodes. To prove our concept, we have electrodeposited 1-μm diameter Pd single wires and small arrays. To demonstrate nanowire sensors, we have electrochemically grown metal (Pd, Au, Pt), metal oxide ($Sb_2O_3$), and conducting polymer (polyaniline) bundled nanowires. Using Pt bundled nanowires surface modified with glucose oxidase, we have demonstrated glucose detection as a demonstration of a biomolecular sensor.

Keywords: Nanowire, Sensor

## 1. INTRODUCTION

There is a great deal of interest in nanostructures for potential applications in such areas as electronics, biochemistry, materials, and medicine. One-dimensional structured materials, such as nanowires and carbon nanotubes (CNTs), are candidate materials for these applications of nanotechnology. Many researchers have developed nanosensors based on Si nanowires or CNTs. For example, Cui *et al.* demonstrated that Si nanowire-based sensors are capable of highly sensitive and selective real-time detection of biomolecules[1]. Li *et al.* reported molecular detection based on electrodeposited copper nanowires grown between nano-gap electrodes[2]. Star *et al.* demonstrated CNT based FET devices[3]. However, these building blocks have some limitations. Existing fabrication methods for CNTs produce mixtures of metallic and semiconducting nanotubes, which make them difficult to use as sensing materials since metallic and semiconducting nanowires will function differently. In addition, surface modification methods for CNTs, which are essential to prepare interfaces for selectively binding a wide range of chemical and biological analytes[1] are not well established. Silicon nanowires are produced by a laser-assisted vapor-liquid-solid growth method[4] or a supercritical fluid solution method[5]. Even though Si nanowires are good sensing materials, they have intrinsic drawbacks of process variability and low throughput. These methods also require that nanowires and nanotubes must be manually aligned and then electrically connected by a post-growth assembly process.

Electrodeposited nanowire sensors can overcome the limitations of both CNTs and Si nanowires due to the relative ease of fabrication and surface modification. A wide range of sensing materials can be deposited by electrodeposition, including metals, alloys, metal oxides, semiconductors, and conducting polymers. Electrodeposition allows a high degree of specificity in location and chemical identity of a deposit, as well as control of thickness[6, 7]. The operating principle of nanowire-based biochemical sensors is the detection of low molecular concentrations by measuring changes in the electrical conductance of nanowires produced by the adsorption or bioreaction of the chemical species. We report an approach to growing nanowires for sensor arrays using standard semiconductor device fabrication techniques. This technique can potentially produce individually addressable nanowire sensor arrays with the capability of sensing multiple chemical species simultaneously. This technique involves electrodeposition to directly grow nanowires between patterned thin film contact electrodes, eliminating expensive and tedious post-growth device assembly. Our proposed nanowire-based sensor will also potentially require low power consumption compare to 100 nW for Pd mesowire arrays[8]. In this work, electrodeposition of single wires and small arrays with wire diameters of 1 μm is demonstrated, and glucose sensors based on surface-modified nanowire bundles are demonstrated. Table 1 compares state-of-the art nanosensors with the nanowire array sensor concept described here.

---

* Minhee.Yun@jpl.nasa.gov; phone 1 818 354-3413; fax 1 818 393-4540

Table 1. The comparison of state-of-the art nanosensor materials

| | CNTs | SNWs | This work |
|---|---|---|---|
| **Materials** | Carbon | Silicon | Metal alloy, metal oxide, conducting polymers |
| **Deposition Techniques** | 1.Arc-discharge method<br>2.Laser assisted<br>3.Chemical vapor deposition | 1.Laser assisted<br>2.Supercritical fluid solution method | 1.Electrochemical method |
| **Manufacturability** | Diffcult | Difficult | Easy |
| **Surface Modification** | Limited | Well-known | Well-known |
| **Functionality** | Single species | Single species | Multiple species |
| | 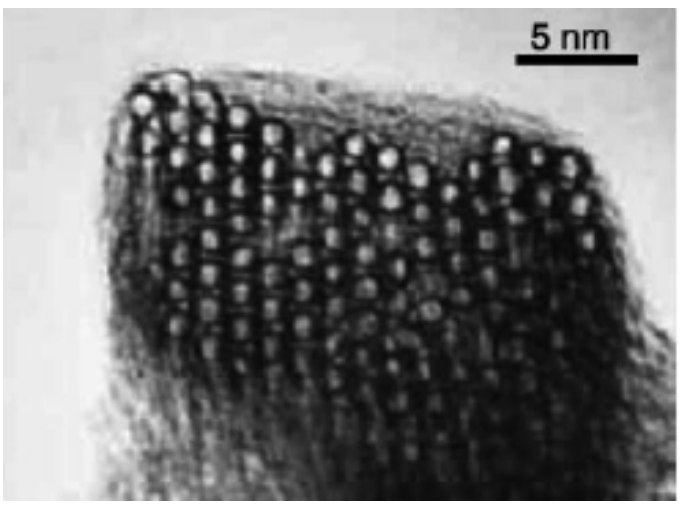  From ref. #9 | 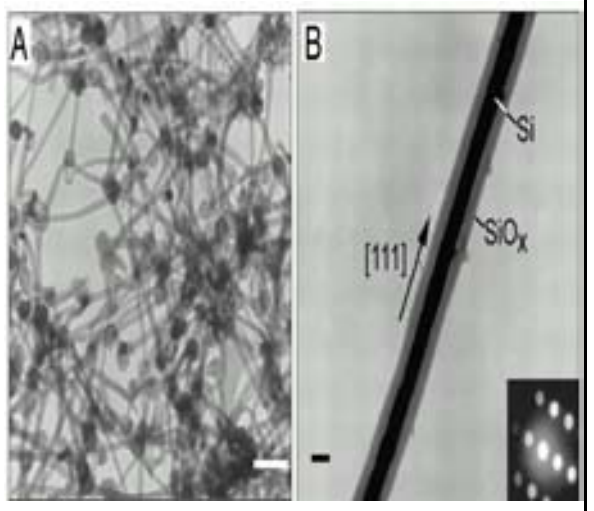  From ref. #7 | 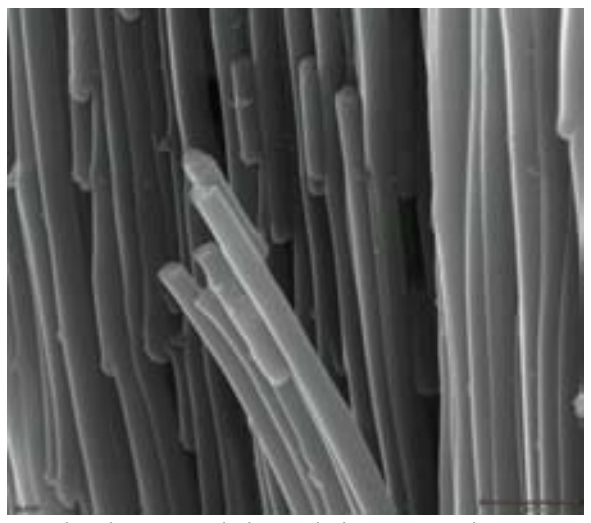 Fabricated in this work |

## 2. FABRICATION

Figure 1 shows a schematic diagram of electrodeposited wires. The processes used in this work, including cleaning, dry etching, low-pressure chemical vapor deposition (LPCVD), lithography, dielectric deposition, e-beam lithography, metallization and electrochemical deposition, are standard semiconductor device fabrication techniques. Si (100) is cleaned with standard RCA cleaning and a 1-μm-thick layer of low stress $Si_3N_4$ insulator is deposited using LPCVD. A 300 nm-thick Ti-Au metal film is deposited and patterned using a liftoff technique to form the contact layer. SiO is then thermally deposited and the electrolyte channel is e-beam patterned and etched using reactive ion etching. Electrochemical deposition is performed by adding one drop of electroplating solution on top of the channel. When an electrical potential is applied between the electrodes, a wire grows from cathode to anode through the channel, which also limits branching dendritic growth.

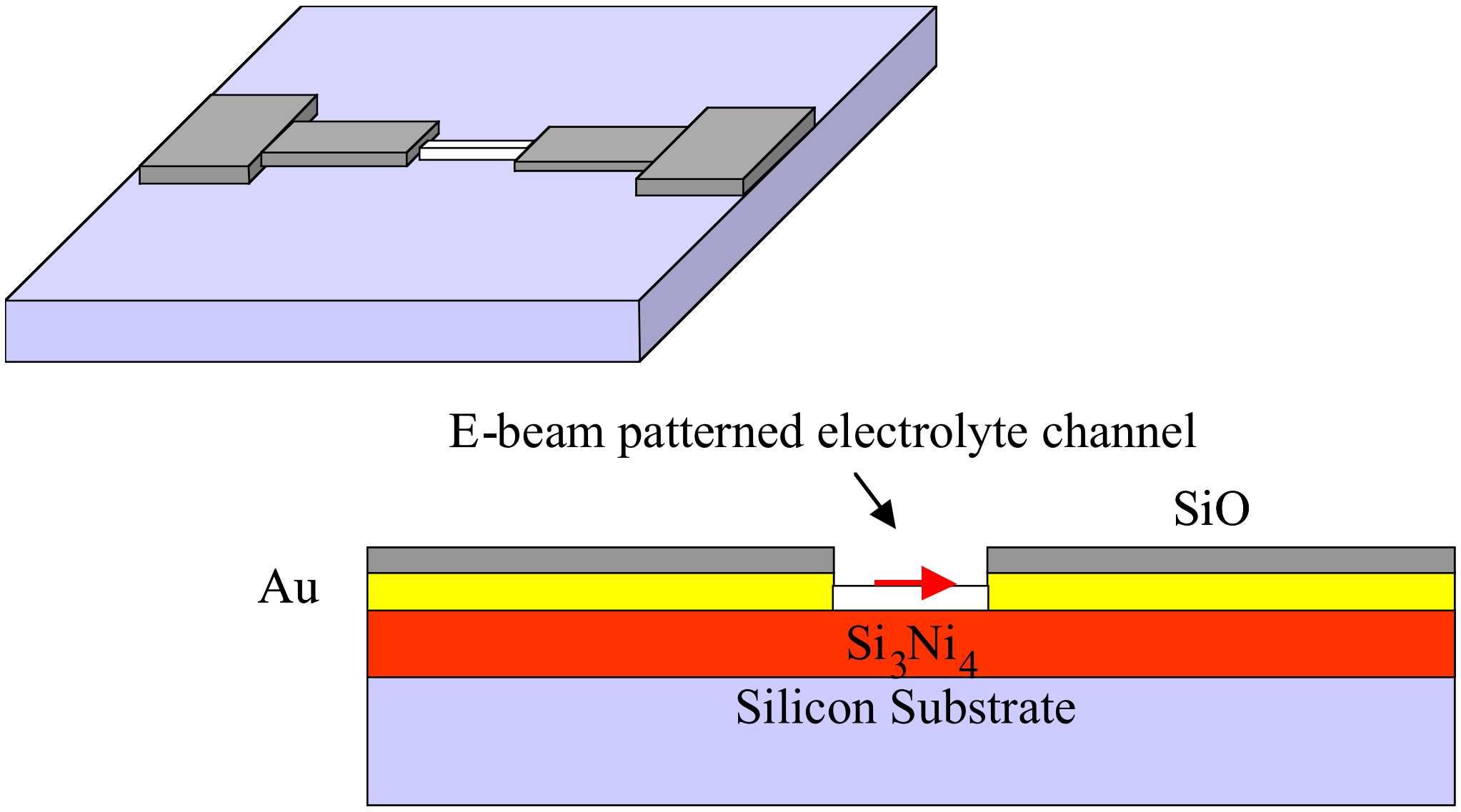

Figure 1. Schematic of electrodeposited wire with contact electrodes.

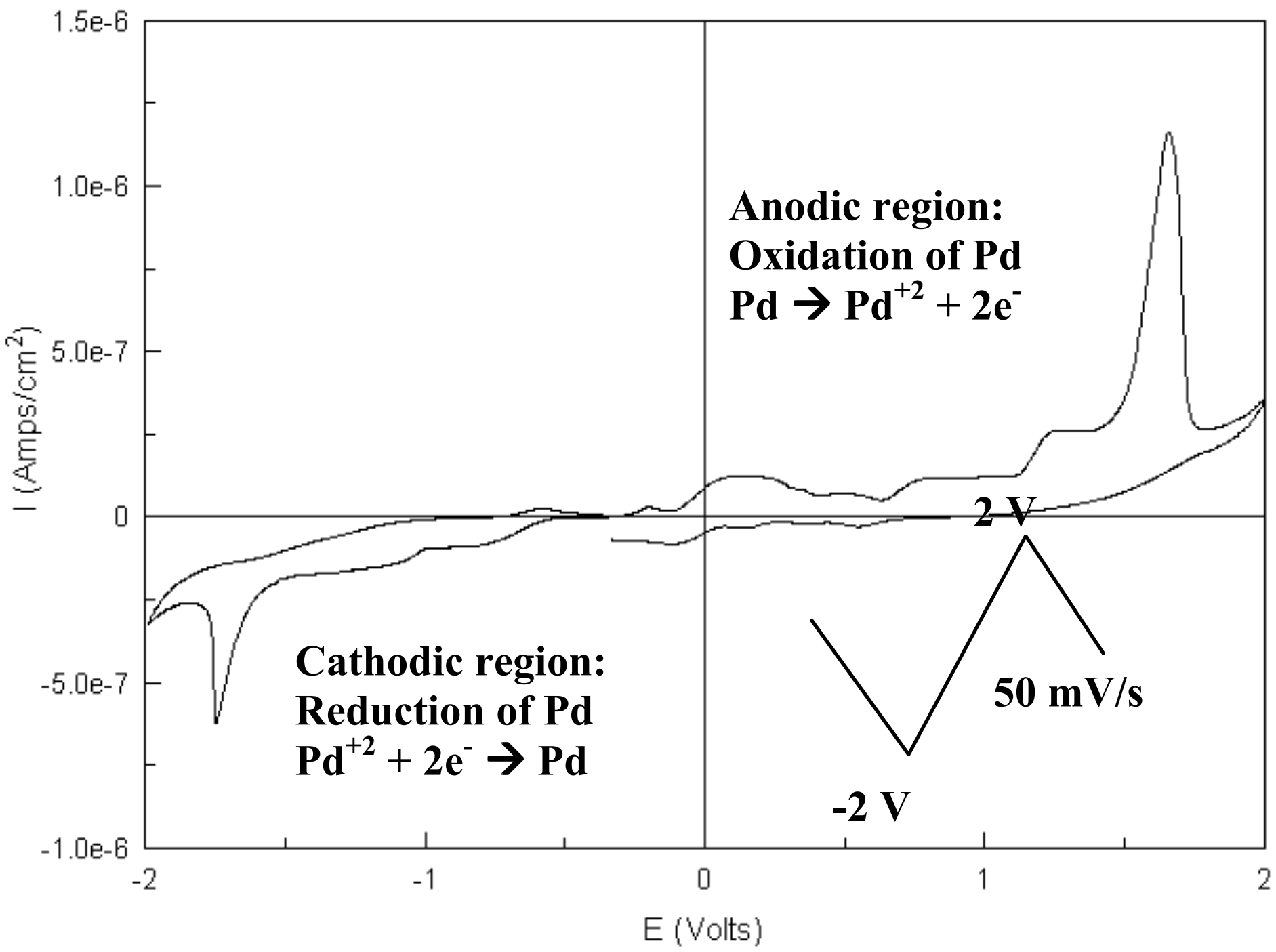

Figure 2. Cyclic voltammogram of Pd electrodeposition in Pd P-salt plating baths: two electrode configuration, scan rate=50 mV/s

The dimensions of the nanowire, including its length and diameter, are predetermined by the width of the nanochannel and the distance between electrodes.

To demonstrate the concept, Pd wires with 1 μm diameter and 3 μm and 7 μm lengths are electrodeposited. Two different electrodeposition solutions (palladium chloride acid bath and palladium p-salt alkaline bath) were initially considered. However, the preliminary experimental results indicate that Pd p-salt solution produces a smoother deposit with minimum dendrite formation at higher cathodic potentials. The Pd electrolyte consists of $Pd(NH_2)_2(NO_2)_2$ (10g/l), and $NH_4NH_2SO_3$ (100 g/l). The pH of the solution is adjusted to 8.0 by addition of $H_3NO_3S$ and NaOH. Figure 2 shows a cyclic voltammogram of Pd p-salt plating solution using a two electrode configuration. The reduction peak of Pd ions to Pd is observed at –1.7 V. When a more negative potential than –1.7 V is applied to the electrode, a significant increase in the current density is observed which is due to $H_2$ gas evolution. A computer controlled EG&G 273 potentiostat/galvanostat is used to grow Pd wires in galvanostatic mode. The applied currents are –10 nA, -20 nA, –100 nA, and –1000 nA and corresponding potentials are monitored.

Figure 3 shows patterned electrode arrays before electrodeposition of wires with 1 μm and 500 nm diameters. The electrolyte channel length is 70 μm and electrode gap is 3 μm, respectively. Figure 4(a) shows cathode potential responses during electrodeposition of a Pd wire at an applied current of –1000 nA grown between electrodes with no electrolyte channel. The cathode potential initially approaches a negative value steeply, followed by a gradual increase in the potential as the Pd wire grows from cathode to anode. When a wire is fully grown and makes contact to the anode, the potential drops to zero and the applied potential is turned off. Lower cathode potentials and shorter deposition times are observed at a higher deposition current as expected due to a higher deposition rate. The 7–μm-long Pd wires were grown within 1500 seconds with –1000 nA. Figure 4(b) shows the changes in electrical resistance between Au electrodes during Pd wire growth at –1000 nA. As expected, the electrical resistance gradually decreases as the Pd wire grows from cathode to anode and reduces the gap. When the Pd wire makes contact to the anode, the measured resistance is less than 100 Ω in the liquid electrolyte. Optical images of electrochemically grown Pd wires between Au electrodes are shown in Fig. 5. The length of the wire is approximately 7 μm and the width is approximately 1 μm. Fig. 5(b) shows double Pd wires directly grown between common Au electrodes.

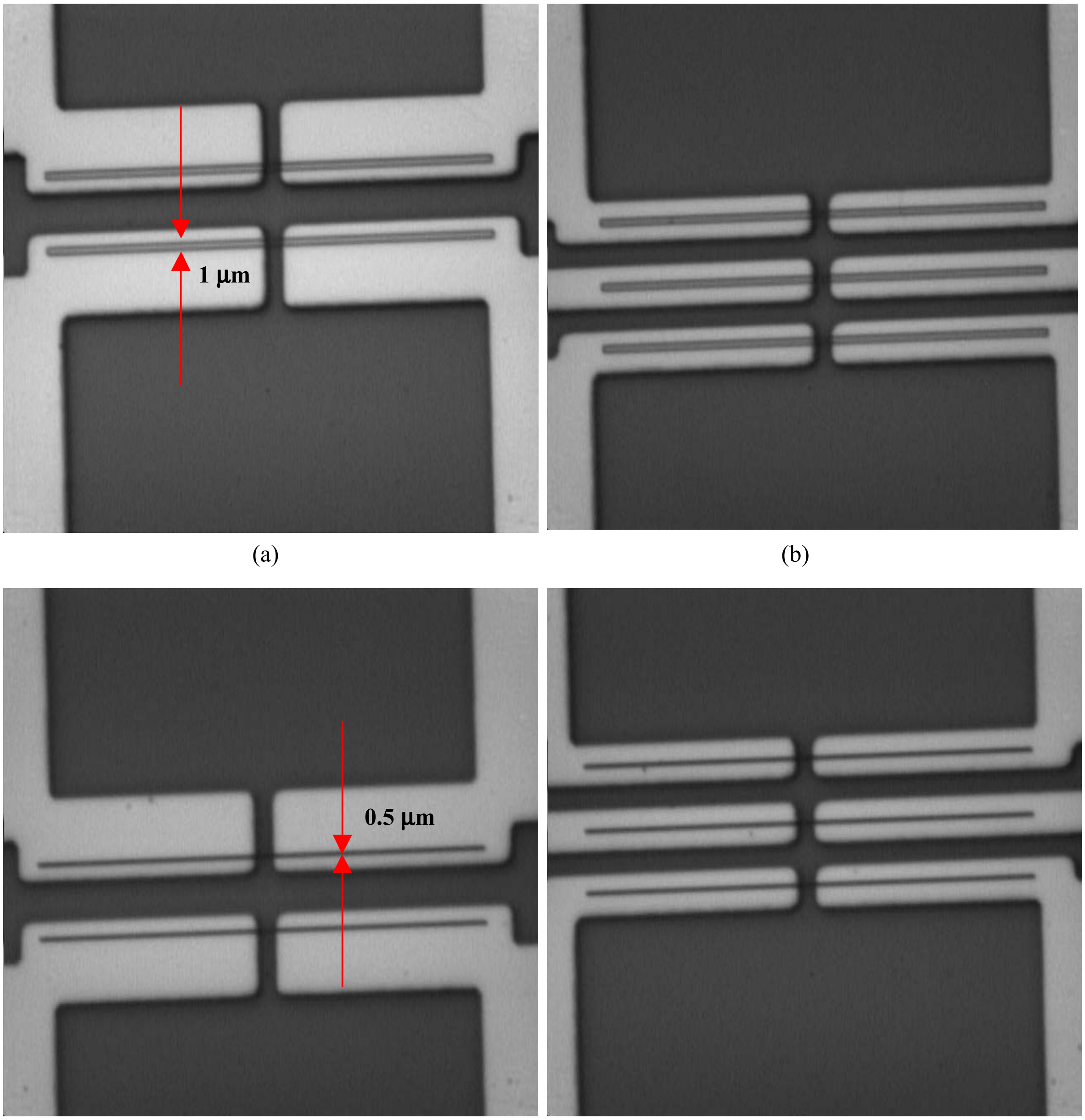

(c) (d)

Figure 3. Electrode arrays with e-beam patterned electrolyte channels before the growth of (a) 1 µm width and 3 µm length wires (2 electrodes), (b) 1µm width and 3 µm length (3 electrodes), (c) 500 nm width and 3 µm length (2 electrodes), and (d) 500 nm width and 3 µm length (3 electrodes).

We have thus successfully demonstrated growth of 1 µm diameter wires and the growth of wires with diameters smaller than 1 µm is currently under investigation. These small arrays are designed for one, two, or three sensing elements. Reducing the width of the e-beam patterned channels, which is currently under investigation, can further reduce the width of electrodeposited wires to a few tens of nm. We are also currently investigating utilizing different electrolytes to fabricate small arrays with wires of different compositions, and hence different chemical sensing capabilities.

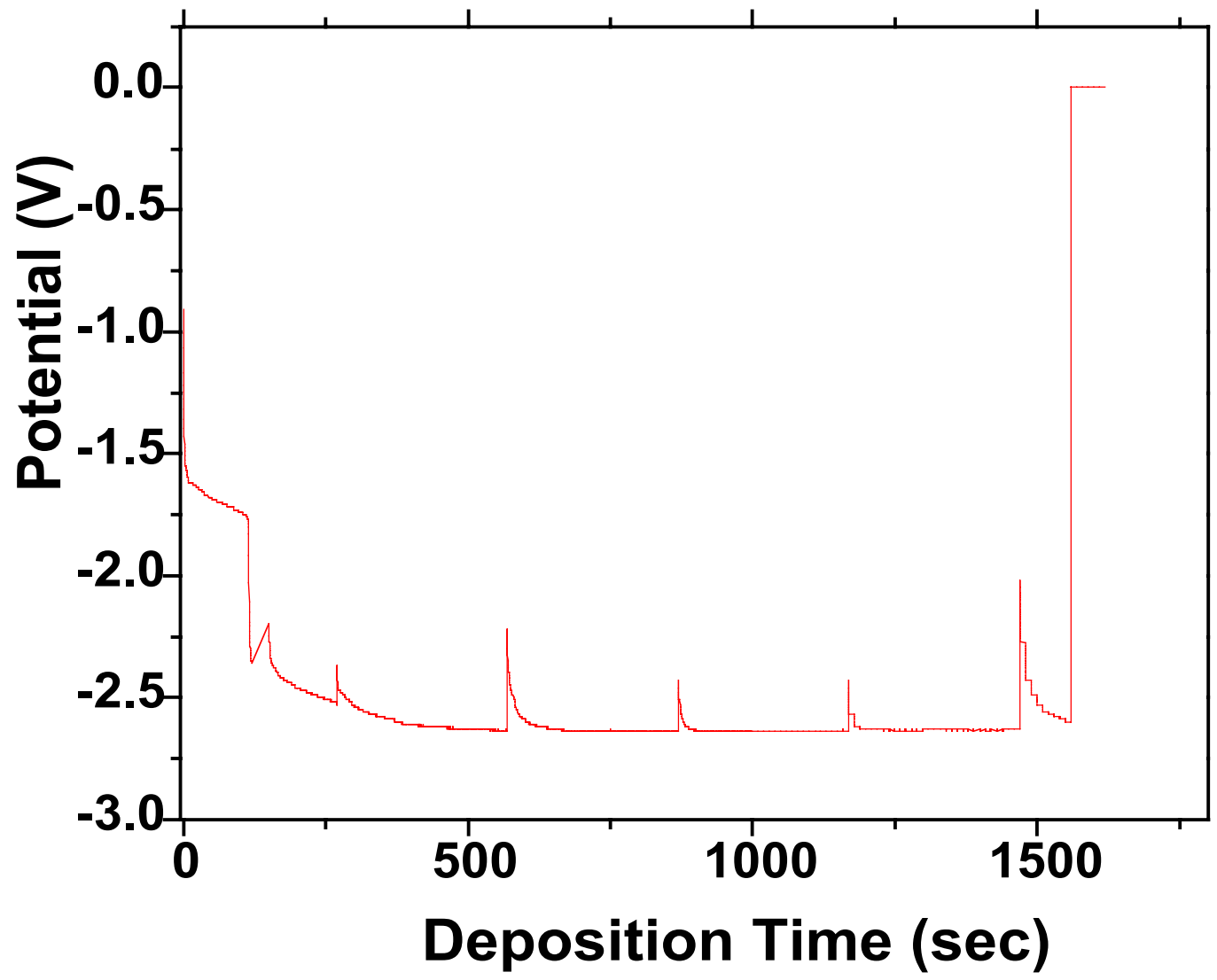

(a) Cathode potential responses as a function of deposition time: The applied current was –1000 nA.

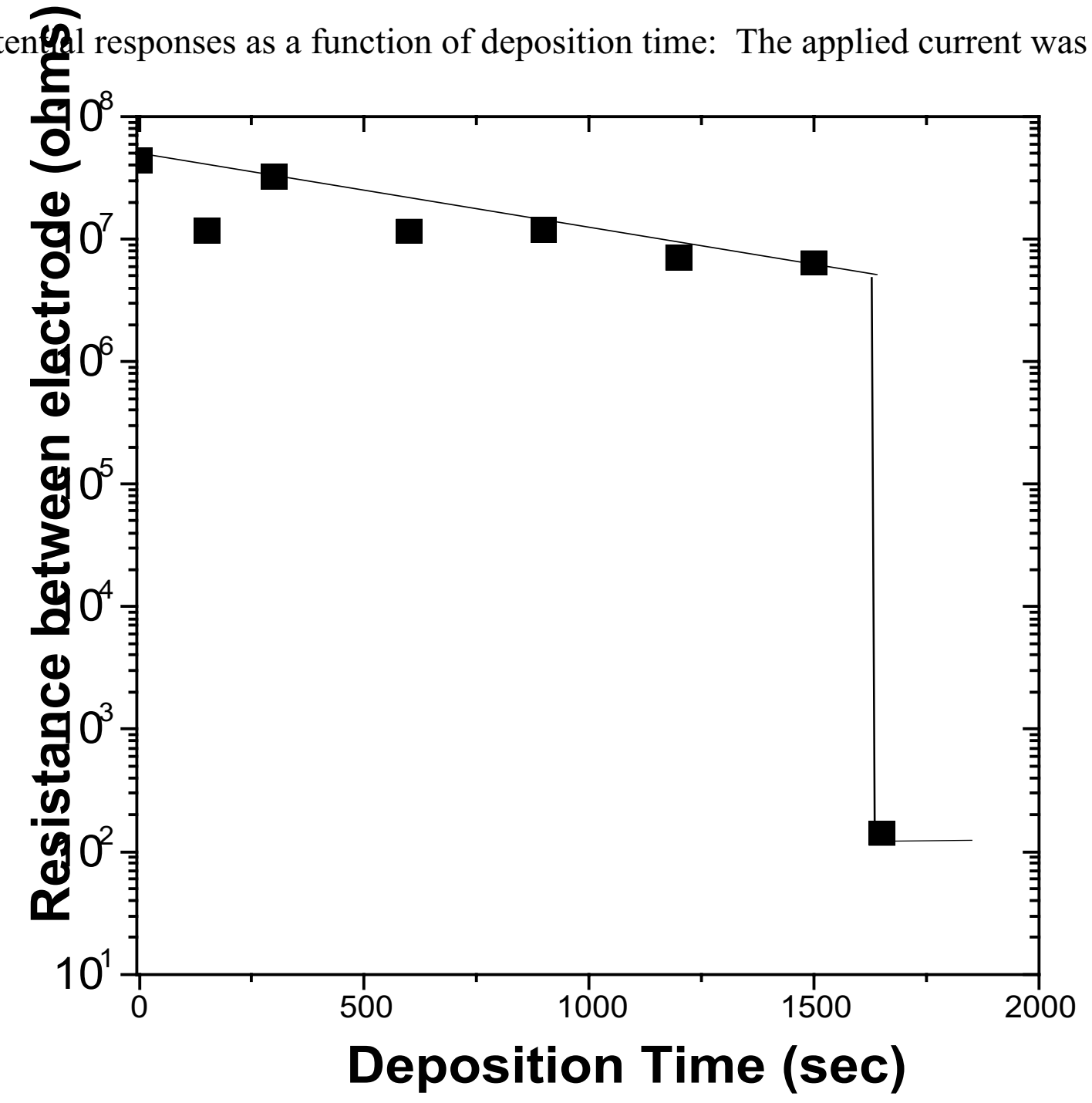

(b) Resistance change between anode and cathode as a function of time: Deposition current was kept at –1000 nA.

Figure 4. Measured electrical properties during deposition: (a) deposition potential versus deposition time and (b) resistance changes versus deposition time between electrodes with no electrolyte channel.

## 3. NANOWIRE SENSORS: DEMONSTRATION AND CONCEPTS

While the ultimate goal is to fabricate individually addressable nanowire sensor arrays, single sensors based on nanowire bundles are also being investigated. This approach allows development of nanowire surface modification techniques and characterization of sensor sensitivity and selectivity in parallel with the development of the nanowire array fabrication.

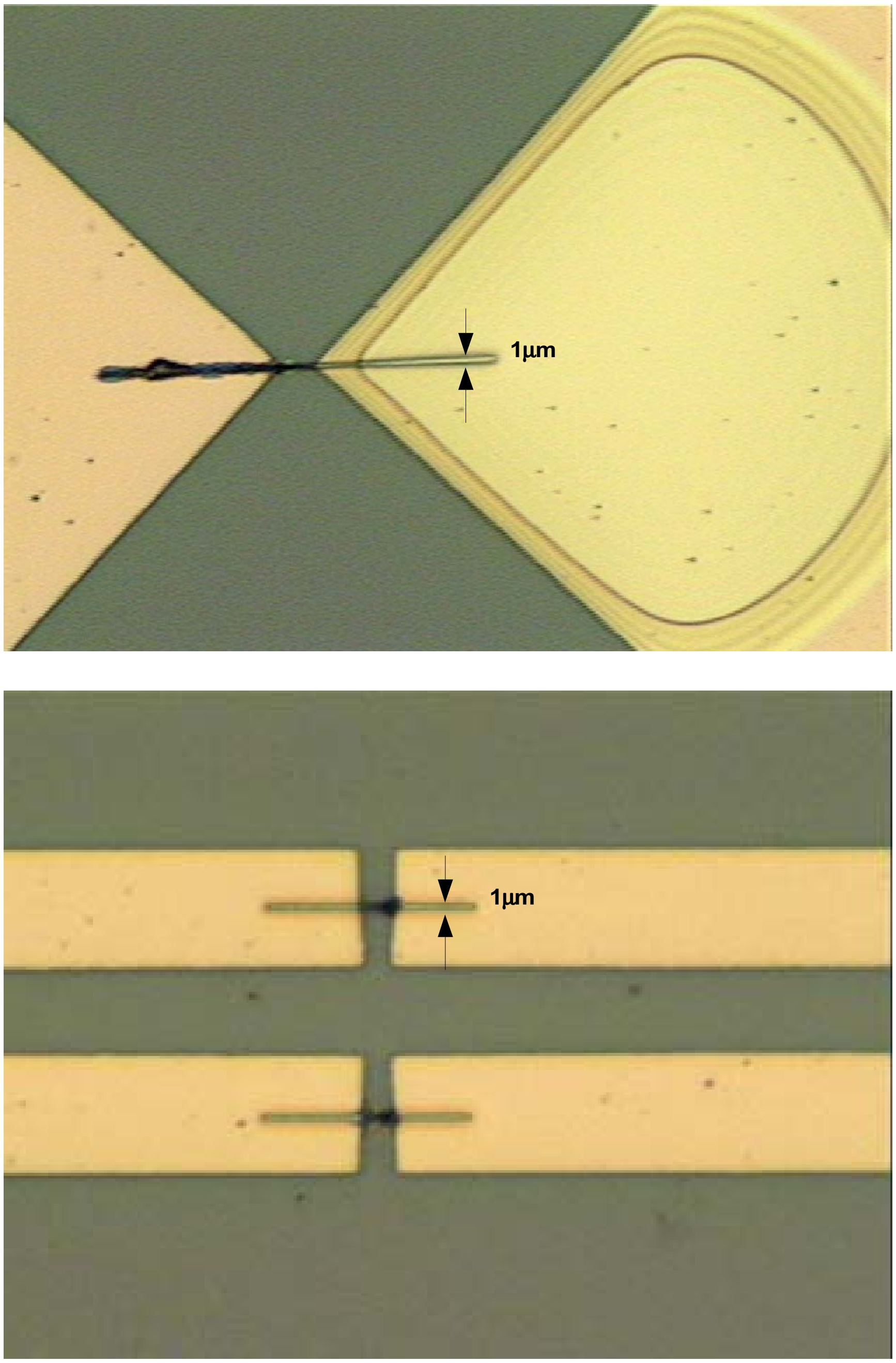

Figure 5. Optimal images of electrodeposited Pd wires grown between electrodes: (A) single wire, and (B) double wires grown between common electrodes.

**Bundled Nanowire Fabrication**

Pt bundled nanowires are fabricated using the electrochemical method. Anodic templates with porosity of 0.43 (Whatman Inc. – 100 and 200 nm diam.) were utilized to electrodeposit Pt nanowires. Prior to electrodeposition, gold was evaporated (200 nm thick) on one side of the alumina template and served as the conductive seed layer. Carbon paste was used to connect the alumina nanotemplate with the seed layer to the silicon substrate. The platinum plating solution consists of 1 g/l $H_2PtCl_6$ + 176.4 g/l $H_2SO_4$. The solution pH was less than 1 and the deposition current density was fixed at 35 mA $cm^{-2}$. The length of the nanowires was controlled by adjusting deposition times. After electrodeposition of nanowires, concentrated KOH or NaOH (20 v/v %) was used to remove the anodized alumina to free the nanowires. Other bundled nanowires (Pd, $Sb/Sb_2O_3$, Au, and polyaniline) have been fabricated using same method but using different electrolyte solutions. For example, Pd bundled nanowires were electrodeposited from 10 g/l $Pd(NH_2)_2(NO_2)_2$ and 100 g/l ammonium sulfamate and $Sb/Sb_2O_3$ bundled nanowires were electrodeposited from 0.03 M potaasium antimonyl tartrate and 0.435 M of sodium tartrate dihydrate. The pH of Pd and $Sb/Sb_2O_3$ plating solutions was adjusted to 8 and 7, respectively. In both case, the applied current density was fixed at 10 mA $cm^{-2}$. Gold and polyaniline were electrodeposited from gold sulfite and aniline-sulfuric acid baths, respectively.

**Biochemical Sensors**

Amperometric biosensors can be created by electronically coupling the appropriate redox enzymes to a metal electrode modified with a self-assembled monolayer (SAM) to facilitate enzyme immobilization and to reject interfering species. Conductometric biosensors are assembled by entrapping the relevant enzymes in conjugated polymer nanowires (e.g., polyaniline) on an electrode. Both approaches entail straightforward synthesis protocols, yet the SAM-based system on Au provides for better interferent rejection while the conductometric system gives substantial signal amplification due to the large change in polymer conductivity in response to small perturbations in its microenvironment. Amperometric glutamate biosensors have been described based on the enzyme glutamate oxidase (gluOX)[10]. Glutamate oxidase, which can be immobilized on the electrode surface by a variety of techniques, catalyzes the oxidation of the amino acid glutamate to α-ketoglutarate and ammonia using oxygen as the electron acceptor.

$$\text{L-glutamate} + O_2 + H_2O \rightarrow \alpha\text{-ketoglutarate} + NH_4^+ + H_2O_2, \qquad \text{Eq. (1)}$$

The hydrogen peroxide generated as a result of the enzyme catalyzed reaction can be oxidized at the electrode surface to give a measurable current that can be correlated to the glutamate concentration.

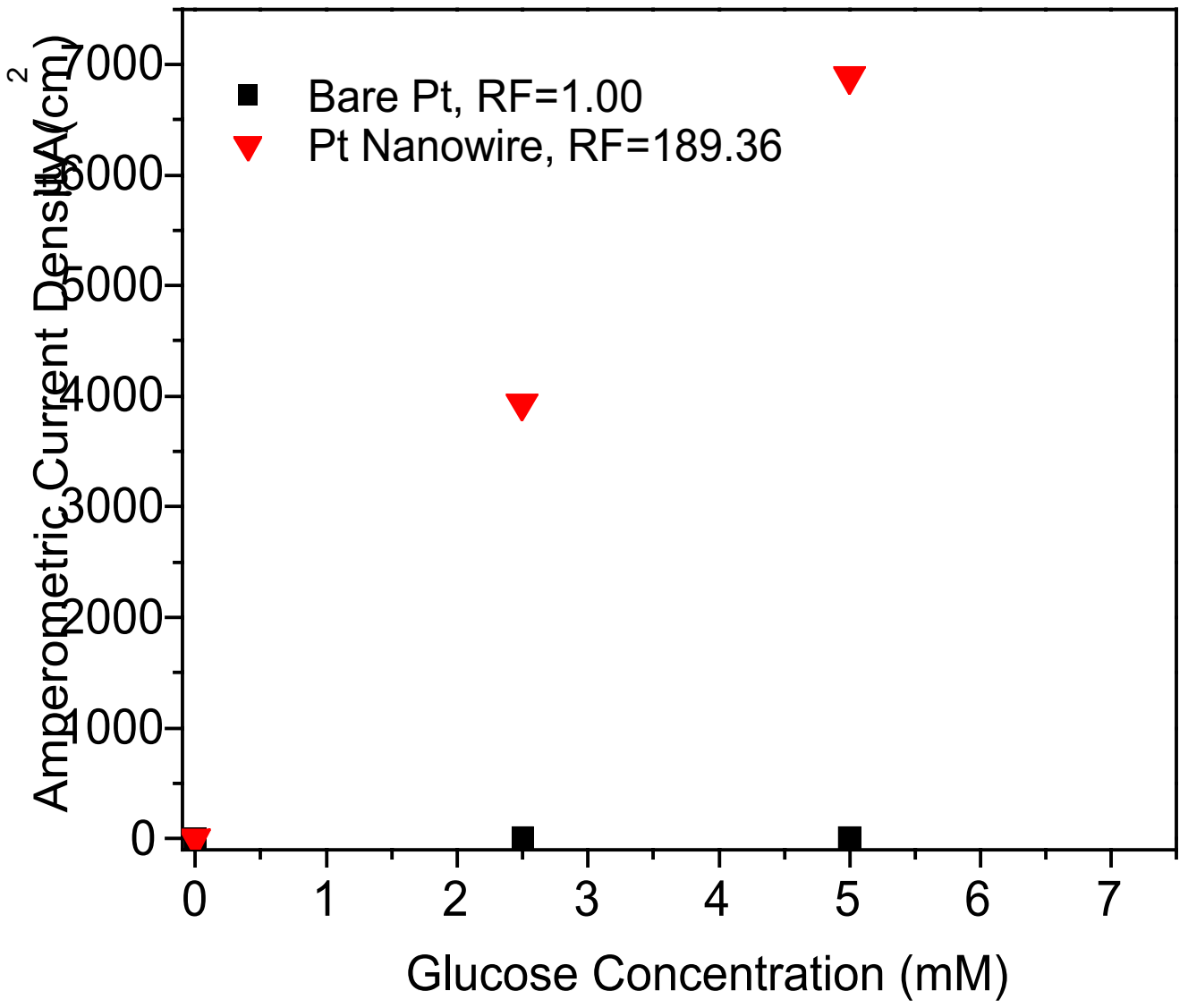

Figure 6. Comparison between amperometric response of glucose sensor constructed on Pt (_) thin film and (_) bundled nanowires. The Pt bundled nanowires have a high roughness factor (RF), i.e. it has a high effective surface area. The current was measured at +0.7 V vs Ag/AgCl at room temperature.

To prove our concept, we have demonstrated a glucose sensor using bundled Pt nanowires because the fundamentals of glucose and glutamate nanowire sensors are similar. An ultrasensitive glucose amperometric sensor was constructed by immobilizing glucose oxidase on Pt bundled nanowires. Glucose oxidase was immobilized by a Galvanostatic process, in which a constant current density of 382 _A/$cm^2$ for 250 seconds was applied. The solution is 2000 U/ml glucose oxidase, 200mM pyrrole using 100mM KCl as the supporting electrolyte. This electrochemically-assisted enzyme immobilization technique not only controls the polymer thickness accurately, but also enables a precise enzyme deposition on small electrodes. Figure 6 compares the amperometric responses of glucose sensors constructed on a Pt thin film and on Pt bundled nanowires. All the measurements were taken at a working potential of 0.7 V vs Ag/AgCl at room temperature. The response to glucose is tremendously improved by three orders of magnitude by using Pt nanowire as sensor material. This is attributed to the ultrahigh surface area of Pt nanowire bundle, which was determined by cyclic voltammogram in 1N $H_2SO_4$ media.

**Gas Sensor**

Nanowire-based gas sensors can potentially exhibit a fast response with a substantially higher sensitivity and selectivity than existing sensors. The basic principle behind nanowire-based gas sensors is the detection of small concentrations by measuring changes in electrical conductance in nanowires produced by the adsorption of the chemical species or by phase changes in nanowires.

In this work, Pd nanowires are being investigated for their capability to sense $H_2$. Pd has low contact resistance and high sensitivity to $H_2$. Favier et al. demonstrated the activity of electrodeposited multiple Pd nanowires prepared on graphite surfaces as $H_2$ sensors and hydrogen-activated switches by applying a constant voltage of 5 mV between Ag contacts and measuring current[8]. The resistance change is caused by a phase change from metallic Pd to $PdH_2$. These Pd nanowire sensors operate at room temperature, have a fast response time (<75 msec), require low power (<100 nW), and are resistant to poisoning by reactive gases, including $O_2$, CO, and $CH_4$. In this work, single Pd wires and bundled nanowires were electrodeposited with controlled width and length from electrolytes based on Pd salts. Demonstrating sensing with single nanowires will be the objective of future work.

**pH Sensors**

The concentration of the $H_3O^+$ is a critical parameter to be measured for monitoring the condition of aqueous biological species, or for predicting the path of chemical reactions. The most widely used solid-state metal oxides used for pH sensing have been potentiometric $Sb_2O_3$ sensors. They have fast response time, "drift-free" behavior, and show good stability in highly acidic and alkaline environments. In contrast to bulk glass pH electrodes, solid-state metal oxide electrodes are also easily microfabricated. We have fabricated $Sb_2O_3$ bundled nanowires by electrodepositing $Sb_2O_3$ coating on prefabricated inert nanowires (e.g. Au or Pt). Fabrication of these $Sb_2O_3$ nanowires into pH sensors is under investigation.

**A Biomedical Sensor** can be fabricated using Au and polyaniline nanowires. We are presently investigating fabricating a sensor to detect the thyroid hormones 3,5,3'-triiodo-L-thyronine (T3) and thyroxine (T4) using bundled Au and polyaniline nanowires.

## 4. SUMMARY

We have developed a fabrication technique that is potentially capable of producing arrays of individually addressable nanowire sensors with controlled dimensions, positions, alignments, and chemical compositions and are in the process of fabricating sensor arrays to detect gases, biochemicals, and hormones. The concept has been demonstrated by growing Pd wires with 1-μm diameters and 7-μm lengths. Reducing the width of the e-beam patterned channels, which is currently under investigation, can further reduce the width of electrodeposited wires to a few tens of nm. It is envisioned that these are the first steps towards producing nanowire sensor arrays capable of simultaneously detecting multiple chemical species. Large-scale arrays may also be possible, e.g. by utilizing integrated large-scale microfluidic networks[11] to control flow of different electrolytes to nanowire growth electrodes. We have successfully demonstrated a biochemical sensor capable of detecting glucose and are investigating a biomedical sensor capable of detecting thyroid hormones using bundled nanowires. The use of single nanowires for gas, biochemical, and biomedical sensor applications is also being investigated.

## ACKNOWLEDGMENTS

This research was performed at the Jet Propulsion Laboratory, California Institute of Technology, under a contract with National Astronautics and Space Administration.